\documentclass[a4paper,11pt]{article}
\usepackage{pos}

\usepackage{xspace}
\usepackage{bbm}

\newcommand{\repoURL}{https://github.com/evanberkowitz/latex-base}

\newcommand{\Figref}[1]{Figure~\ref{fig:#1}\xspace}
\renewcommand{\eqref}[1]{(\ref{eq:#1})\xspace}

\newcommand{\Reference}[1]{Reference~\cite{#1}}

\newcommand{\goesto}{\ensuremath{\rightarrow}}
\newcommand{\infinity}{\infty}

\renewcommand{\mod}[1]{\ensuremath{\ \left(\text{mod }#1\right)}}

\newcommand{\adjoint}{\ensuremath{{}^{\dagger}}}

\newcommand{\D}{\ensuremath{\mathcal{D}}\xspace}
\renewcommand{\O}{\ensuremath{\mathcal{O}}\xspace}
\newcommand{\RR}{\ensuremath{\mathbb{R}}\xspace}
\newcommand{\Z}{\ensuremath{\mathcal{Z}}\xspace}
\newcommand{\ZZ}{\ensuremath{\mathbb{Z}}\xspace}

\newcommand{\supervillain}{\texttt{supervillain}\xspace}

\let\builtinLaTeX\LaTeX
\def\LaTeX{\builtinLaTeX\xspace}

\usepackage{graphicx}
\usepackage{placeins}

\title{Generalized BKT Transitions and Persistent Order on the Lattice}

\author*[a]{Evan Berkowitz}
\author[b]{Seth Buesing}
\author[c]{Shi Chen}
\author[c]{Aleksey Cherman}
\author[d]{\\Srimoyee Sen}

\affiliation[a]{Institute for Advanced Simulation, J\"{u}lich Supercomputing Centre, and the Center for Advanced Simulation and Analytics (CASA)\\
  Forschungszentrum J\"{u}lich \\
  Wilhelm-Johnen-Stra{\ss}e, 50825 J\"{u}lich, Germany
  }
\affiliation[b]{
	Macalester College, 1600 Grand Ave, St.\@ Paul, MN 55105, USA and \\
	Department of Applied Mathematics and Theoretical Physics \\
	University of Cambridge, Cambridge CB3 0WA, UK
}
\affiliation[c]{School of Physics and Astronomy, 
University of Minnesota, Minneapolis, MN 55455, USA}
\affiliation[d]{Department of Physics and Astronomy, 
Iowa State University, Ames, IA 50011, USA}

\emailAdd{e.berkowitz@fz-juelich.de}

\abstract{
	The BKT transition in low-dimensional systems with a U(1) global symmetry separates a gapless conformal phase from a trivially gapped, disordered phase, and is driven by vortex proliferation.
	Recent developments in modified Villain discretizations provide a class of lattice models which have a $\mathbb{Z}_W$ global symmetry that counts vortices mod W, mixed 't Hooft anomalies, and persistent order even at finite lattice spacing.
	While there is no fully-disordered phase (except in the original BKT limit $W=1$) there is still a phase boundary which separates gapped ordered phases from gapless phases.
	I'll describe a numerical Monte Carlo exploration of these phenomena.
}

\FullConference{The 41st International Symposium on Lattice Field Theory (LATTICE2024)\\
 28 July - 3 August 2024\\
Liverpool, UK\\}

\begin{document}
\maketitle

\section{Introduction}\label{sec:introduction}

Recent developments~\cite{Berkowitz:2023pnz,Morikawa:2024zyd} give a direct lattice formulation of a 2D chiral gauge theory, building on the modified Villain construction.
The restriction to 2D comes from leveraging bosonization: rather than discretize the Dirac operator $D$, the construction first trades the continuum fermion determinant for a bosonic path integral and carefully discretizes the bosonic action using the modified Villain formulation~\cite{Sulejmanpasic:2019ytl,Gorantla:2021svj}.
This maintains the continuum symmetries and 't Hooft anomalies exactly on the lattice.
The modified Villain constructions of 1- and 2-flavor QED and the 3450 chiral gauge theory in \Reference{Berkowitz:2023pnz} all follow this strategy.

Critical to these constructions is a careful discretization of the compact boson.
The compact boson in 1+1D
\begin{align}
	\label{eq:bosonization}
	\Z &= \int \D\varphi \exp\left\{- \int d^2x \frac{1}{8\pi} (d\varphi)^2\right\}
	&
	\varphi & \sim \varphi + 2\pi
\end{align}
is dual to a free fermion~\cite{coleman1975charge,coleman1976more}, and has two interesting global $U(1)$ symmetries.
The first is the shift and the second the winding symmetry,
\begin{align}
	\text{shift }\qquad &\qquad U(1)_S
	&&
	\varphi \rightarrow \varphi + \varepsilon
	&
	J^s_\mu &= \frac{i}{4\pi} \partial_\mu \varphi
	\\
	\text{winding }\qquad& \qquad U(1)_W
	&&
	\text{topological, not Noetherian}
	&
	J^w_\mu &= \frac{1}{2\pi} \epsilon_{\mu\nu} \partial^\nu \varphi,
\end{align}
which correspond to the vector and axial currents on the fermionic side.
The first is conserved by the equations of motion of $\varphi$, while the second is conserved \emph{as long as partial derivatives commute}.
That caveat amounts to the condition that $\varphi$ harbors no vortices, because partial derivatives fail to commute near vortices, as parallel transport around a vortex accumulates a net winding number.
These $U(1)$s have a mixed 't Hooft anomaly.

The vortices play a crucial role.
If we relax the constraint that there are no vortices (so that the topological current is not conserved), we are left with the XY model, which undergoes the BKT transition when vortices are entropically favored and proliferate, destroying a CFT phase and yielding a trivially gapped phase.
In contrast, when the vortices are prohibited and the winding $U(1)$ symmetry is exact and the 't Hooft anomaly guarantees we cannot get a trivial phase.

While also preparing us for more interesting field theories in the future, studying the compact boson alone still offers interesting physics.
In fact, we will see that on the lattice it is natural to study a family of models that interpolate between the conventional XY model and a discretization of the compact boson \eqref{bosonization} that exactly preserves its global symmetries and anomalies at finite lattice spacing, even with a finite number of degrees of freedom.

In the XY model the $U(1)_W$ symmetry is completely broken.
We will consider lattice models that instead maintain a $\ZZ_W$ subgroup, where $W$ is an integer parameter of our choosing.
The shift $U(1)_S$ and winding $\ZZ_W$ continue to share a mixed 't Hooft anomaly, so that the theory cannot support a trivially gapped phase.
However, the models still undergo a generalized BKT-like vortex-condensation transition, as we will see.

The modified Villain construction allows us to maintain these 't Hooft anomalies at $W>1$ exactly on the lattice.
The trivially gapped phase is replaced by a gapped phase where $\ZZ_W$ is spontaneously broken.
The maintenence of the anomalies at finite spacing illustrates that the common idea that anomalies require an infinite number of degrees of freedom and must emerge only in the continuum limit is lore rather than law.

\section{The Villainous Compact Boson}\label{sec:compact boson}

\begin{figure}
	\centering
	\raisebox{-0.5\height}{\includegraphics[width=0.48\textwidth]{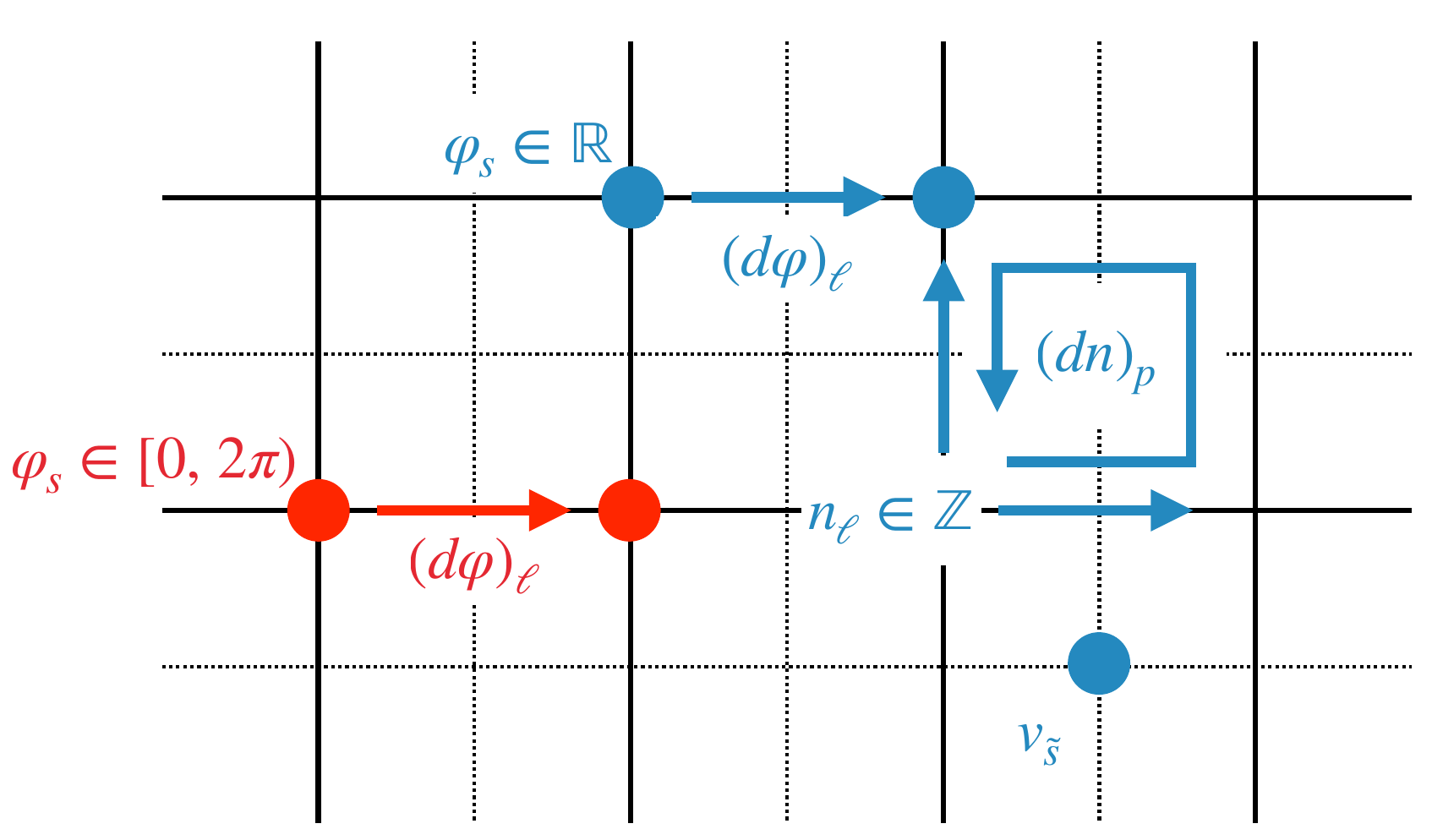}}
	\raisebox{-0.5\height}{\includegraphics[width=0.48\textwidth]{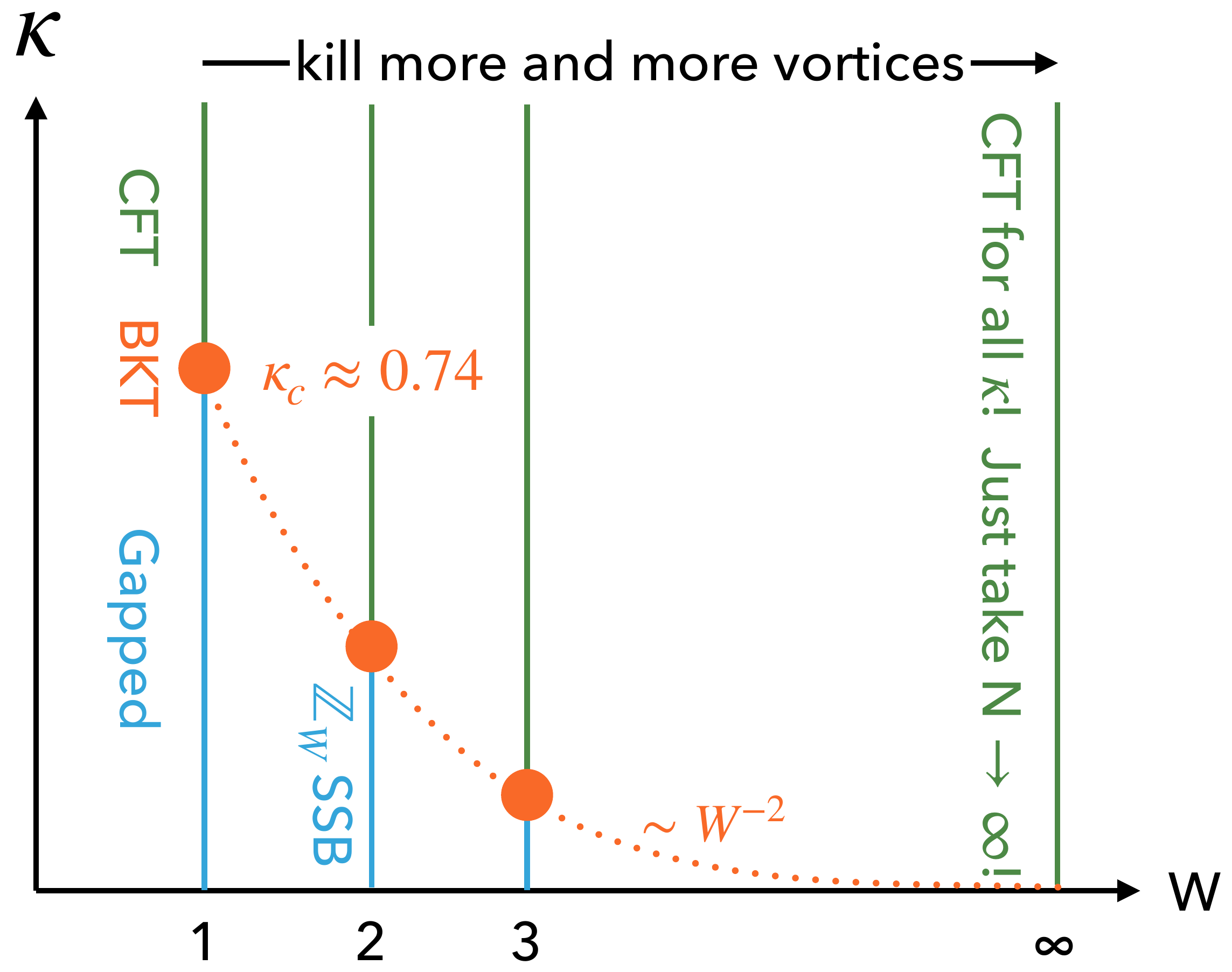}}
	\caption{
		LEFT:
		In the (red) standard Wilsonian construction the XY model comprises $\varphi\in[0,2\pi)$ on sites and the finite difference $d\varphi$ on links of the (solid black) primary lattice.
		In the (blue) Villain construction the model comprises $\varphi\in\RR$ on the sites with the finite difference $d\varphi$ and discrete gauge field $n$ on the links.
		Letting $dn$ on plaquettes be the oriented sum yields a differencing scheme which enjoys $d^2=0$.
		The Lagrange multiplier $v$ lives on sites of the (dotted black) dual lattice, which can be matched with the plaquettes using the Hodge star $\star$.
		RIGHT:
		Increasing $W$ kills more and more vortices, only allowing those $0\mod{W}$, and that allows the CFT (green) to survive at lower couplings $\kappa$.
		In fact, analytic calculations show that $W>1$ undergoes a transition at $\kappa_c/W^2$ where $\kappa_c$ is the critical $W=1$ (BKT) coupling~\cite{Berkowitz:2024bkt}.
		The blue phase is gapped when $W=1$ but ordered to support the 't Hooft anomaly when $W>1$.
		We check this expectation numerically in the top right panel of \Figref{results}.
	}
	\label{fig:lattice}
\end{figure}

To understand modified Villain formulation it is convenient to contrast its formulation of the XY model with the standard Wilsonian formulation.
On an $N \times N$ two-dimensional square lattice with sites $s$, links $\ell$ and plaquettes $p$ the standard formulation is
\begin{align}
	S_{\text{Wilson}} &= \frac{\kappa}{2} \sum_\ell \left[1-\cos(d\varphi)_\ell\right]
	&
	\varphi_s &\in [0,2\pi)
	&&
\end{align}
where $(d\varphi)_\ell$ is the finite difference between $\varphi$ on two neighboring sites.
This action is invariant under the global $U(1)_S$ symmetry $\varphi \goesto \varphi + c$ and the restricted range makes the boson compact.
In contrast in the Villain formulation~\cite{Villain:1974ir,Sulejmanpasic:2019ytl} introduces a discrete gauge field $n$ which lives on links, but otherwise directly transcribes the bosonized action~\eqref{bosonization}
\begin{align}
	S_{\text{Villain}} &= \frac{\kappa}{2} \sum_\ell \left[ (d\varphi)_\ell - 2\pi n_\ell \right]^2
	&
	\varphi_s &\in \RR
	&
	n_\ell & \in \ZZ
\end{align}
and achieves the $2\pi$ periodicity by enforcing the discrete gauge symmetry
\begin{align}
	\varphi_s	&\goesto \varphi_s + 2\pi q_s
	&
	n_\ell	&\goesto n_\ell + (dq)_\ell
	&
	q_s &\in \ZZ
\end{align}
which leaves $(d\varphi - 2\pi n)_\ell$ invariant and leverages the fact that $U(1) \cong \RR/2\pi\ZZ$.
We show these ingredients on the lattice in the left panel of \Figref{lattice}.
Dialing the coupling $\kappa$ amounts to adjusting Thirring terms in the fermionic theory; one particular value is dual to the free fermion.

In the original formulation the exponential of the line element is in the action, as in typical Wilsonian constructions where the group elements enter the action, while in the Villain construction the action uses the algebra elements themselves.
One significant advantage of the Villain construction is that it is quadratic, which lends itself to exact lattice dualities, such as $T$-duality via Poisson resummation to a worldline formulation.

But, the most important difference is the ease with which we can define the vorticity on a plaquette.
In the Wilsonian action one must invent some convention about when to mod by $2\pi$, while in the Villain action we have the obvious analog of the continuum definition of the winding number around some area $w_A$ and always get an integer,
\begin{align}
	\text{continuum}&& w_A &= \frac{1}{2\pi} \oint_{\partial A} \omega &&= \frac{1}{2\pi} \int_A d\omega
	\\
	\text{lattice}&& w_A &= \frac{1}{2\pi} \sum_{\ell\in\partial A} (d\varphi - 2\pi n)_\ell &&= \frac{1}{2\pi} \sum_{p\in A} [d(d\varphi - 2\pi n)]_p = - \sum_{p\in A} (dn)_p \in \ZZ,
\end{align}
using a finite difference which satisfies $d^2=0$.
We emphasize that since $n\in\ZZ$ the Villain winding number is automatically an integer.
A single plaquette can carry any integer winding, while the usual construction typically ascribes winding of $-1$, $0$, or $+1$ on a plaquette, building larger vortices from many plaquettes.

Both actions above yield the same physics of the XY model, with dynamical vortices, the classic BKT transition, and a gapped phase for small $\kappa$.
But we can remove vortices from the Villain construction much more easily by path-integrating over an integer-valued field $v$ which lives on dual sites $\tilde{s}$ and using it as a Lagrange multiplier field to kill all winding (mod $W\in \ZZ$) vortices,
\begin{align}
	S_W &= S_{\text{Villain}} + \sum_p 2\pi i v_{\star p} (dn)_p / W
	\label{eq:villain w}
\end{align}
since the path integration over $v$ yields constraint that $dn_p \equiv 0 \mod{W}$ on every plaquette.
This action has a $\ZZ_W$ winding symmetry,
\begin{align}
	v &\goesto v + z
	&
	z & \in \ZZ
\end{align}
which can be promoted to the full $U(1)_W$, removing all vortices, by letting $v, z \in \RR$ and $W=2\pi$.

When $W=1$ the constraint is trivial and any assignment of $n$s is a valid configuration, but when $W>1$ we keep only vortices of winding $0\mod{W}$.
This is easy to arrange because it is a constraint plaquette-by-plaquette; in the Wilsonian construction we would have to formulate a much more complicated constraint.

In the right panel of \Figref{lattice} we show the expected ($W$, $\kappa$) phase diagram of these models.
When the coupling $\kappa$ is large one gets a CFT (green).
At some intermediate critical coupling (orange) is an infinite-order transition which when $W=1$ is the BKT transition.
Analytic arguments show that if the unconstrained $W=1$ model undergoes the BKT transition at $\kappa_c$, the $W>1$ model undergoes a transition at $\kappa_c / W^2$~\cite{Berkowitz:2024bkt}.
Below the transition the BKT case gives a trivial gapped phase, but when $W>1$ the CFT transitions to an ordered phase to satisfy the lattice 't Hooft anomaly.
When $W=\infty$ we get the compact boson and land on a CFT in the infinite-volume limit \emph{for any coupling whatsoever, with no fine-tuning}.

To detect the transition we will measure two-point correlators.
In a generic CFT a two-point correlator has a simple form,
\begin{align}
	\left\langle \O\adjoint_x \O_y \right\rangle
	&=
	\frac{\text{some OPE coefficient}}{|x-y|^{2\Delta}},
\end{align}
decaying with critical exponent $\Delta$.
While it is conventional to normalize the operators $\O$ so that the numerator is 1, it is UV-sensitive and we will therefore not fix the normalization in our numerical computations.
In our particular case, the CFT is the $c=1$ compact scalar, and the scaling dimensions of all local operators are known.
This information can be used to determine the scaling dimension $\Delta$ of the spin and vortex operators at the critical coupling $\kappa_c(W)$
\begin{align}
	\text{Spin operator}\qquad \O_s &= e^{i\varphi_s}
	&
	\Delta_S(\kappa_c)	&= W^2/8
	\\
	\text{Vortex operator}\qquad \O_p &= e^{2\pi i v_p / W}
	&
	\Delta_V(\kappa_c)	&= 2/W^2,
\end{align}
even though the CFT data does not determine $\kappa_c$ itself, which must be measured numerically.
It is standard to compute the susceptibility, or volume-average, of two-point functions and study its scaling as the size of the lattice $N\goesto\infty$.
However, when $W\geq3$ the spin susceptibility no longer has a long-distance divergence, and converges, preventing the usual finite-size scaling analysis.
Instead we compute the spin and vortex \emph{critical moments},
\begin{align}
	S(x) &= \left\langle e^{i(\varphi_0 - \varphi_x)} \right\rangle
	&
	C_S &= \frac{1}{N^2} \int d^2x\; x^{2\Delta_S(\kappa_c)} S(x)
	\label{eq:spin critical moment}
	\\
	V(\tilde{x}) &= \left\langle e^{2\pi i(v_{\tilde{0}} - v_{\tilde{x}})/W} \right\rangle
	&
	C_V &= \frac{1}{N^2} \int d^2\tilde{x}\; \tilde{x}^{2\Delta_V(\kappa_c)} V(\tilde{x})
	\label{eq:vortex critical moment}
\end{align}
which are engineered to go a constant number as $N\goesto\infty$ at the critical coupling.
Just above $\kappa_c$, on the CFT side, $V \sim 1/x^{2\Delta_V(\kappa)}$, the $C_V$ integrand will look like $1/x^{\text{small power}}$, the integral will go like $N^{2-\text{small}}$, and the $N^{-2}$ normalization will send $C_V\goesto 0$ as $N\goesto\infty$~\cite{supervillain-131}.
In contrast, the integrand of the spin critical moment will scale like $\sim x^{\text{positive power}}$ and the normalization will not be enough to make $C_S$ converge as $N\goesto\infty$; it will diverge.
The story reverses just below the transition: $C_V$ will diverge, indicating a spontaneous breaking of the winding $\ZZ_W$, and $C_S$ will vanish with $N\goesto\infty$.

\section{Results}\label{sec:results}

We perform Markov-chain Monte Carlo simulations at $W\in\{1,2,3\}$ for a variety of couplings $\kappa$ using \supervillain~\cite{supervillain}, which we are developing openly on GitHub with documentation on Read the Docs~\cite{supervillain-docs}.
We can compare the $W=1$ XY-model~\cite{Kosterlitz:1974sm} results to \Reference{Janke:1993va}, which found $\kappa_c\approx0.74$.
The obvious reading of the $W$-modified Villain action \eqref{villain w} has a horrible sign problem introduced by the winding constraint.
However, we can trade the horrible sign problem for the constraint that $dn\equiv 0\mod{W}$ and design an ergodic sampling algorithm which never changes $dn \mod{W}$.
The local fluctuations $\varphi$ can be updated holding $n$ fixed, and the fluctuations $n$ can then be updated holding $\varphi$ fixed; the updates to $n$ require care to maintain the constraint.
We can propose changes to a link $n$ by $W$, propose `exact' changes, updating $n$ by $dz$, or updates around a cycle of the torus.
Most effective is the worm algorithm~\cite{supervillain-villain-worm}, which also allows us to measure $V(\tilde{x})$ \eqref{vortex critical moment} on the fly.

When the coupling $\kappa$ is large, individual updates in the Villain formulation are likely to cause big changes in action and get rejected.
We also perform calculations in the worldline formulation, the lattice-exact Poisson-resummed formulation,
\begin{align}
	\Z &= \int \D{m}\; \D{v}\; e^{-S[m, v]} [\delta m = 0]	&	m, v &\in \ZZ
	\\
	S[m, v] &= \frac{1}{2\kappa} \sum_\ell \left(m - \delta v/W\right)^2_\ell + \frac{(\text{\# links})}{2} \ln 2\pi \kappa - (\text{\# sites}) \ln 2\pi
\end{align}
where $\delta$ is the lattice divergence, we include the constants in the action to exactly match the partition function in the Villain framing, and the degrees of freedom are all integer, but with the constraint that the $m$ is divergenceless everywhere.
This formulation can also be updated carefully to maintain the constraint, and a worm~\cite{supervillain-worldline-worm} allows us to measure $S(x)$ \eqref{spin critical moment} on the fly.
Because it has $\kappa$ in the denominator, this formulation gets easier to sample at strong coupling and more difficult at weak coupling.
We have checked that worldline and Villain computations produce the same action, internal energy density, squared internal energy density, and spin- and vortex- two-point functions $S$ and $V$ at intermediate $\kappa$ where both formulations get reasonable acceptance rates.

\begin{figure}[h]
	\begin{center}
	\includegraphics[width=0.49\textwidth]{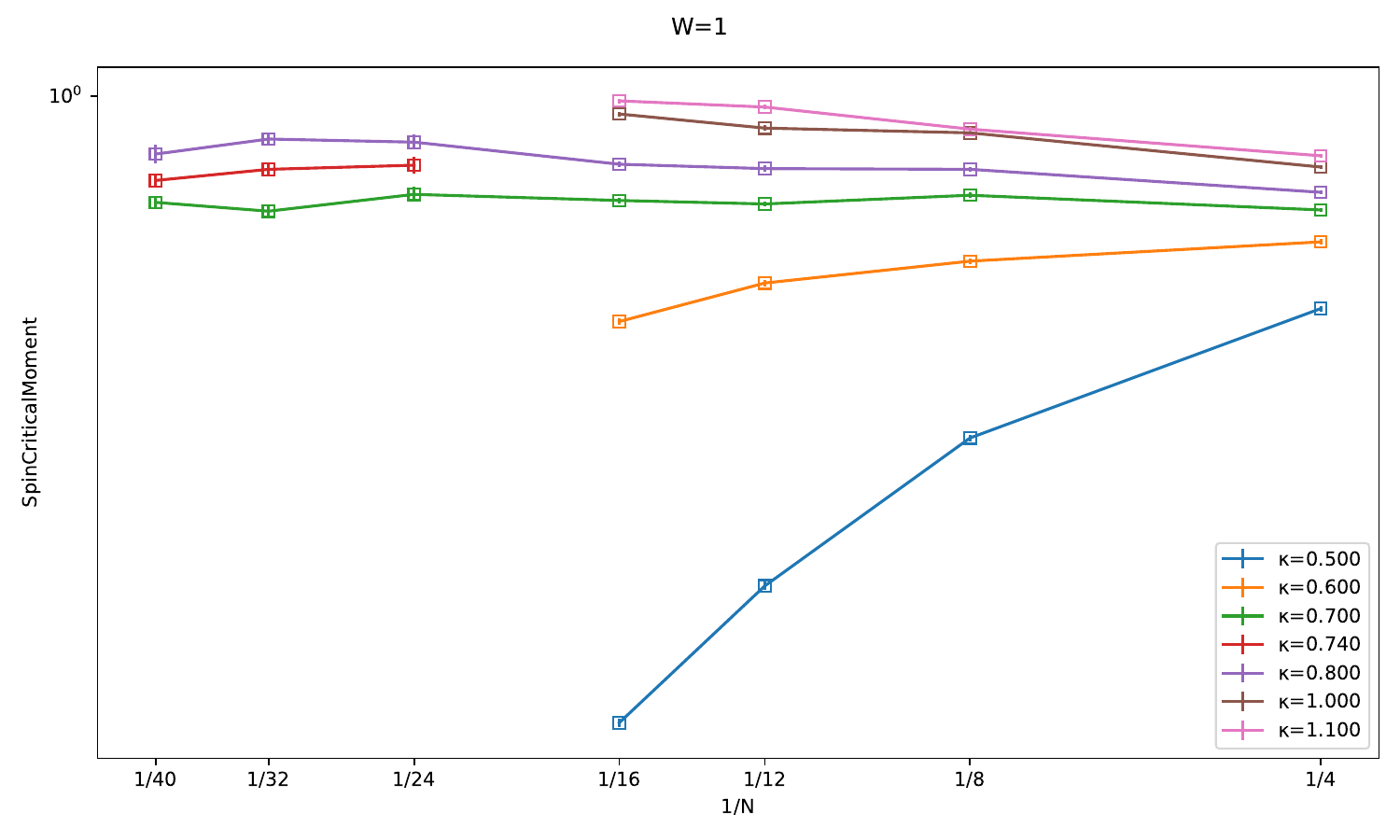}
	\includegraphics[width=0.49\textwidth]{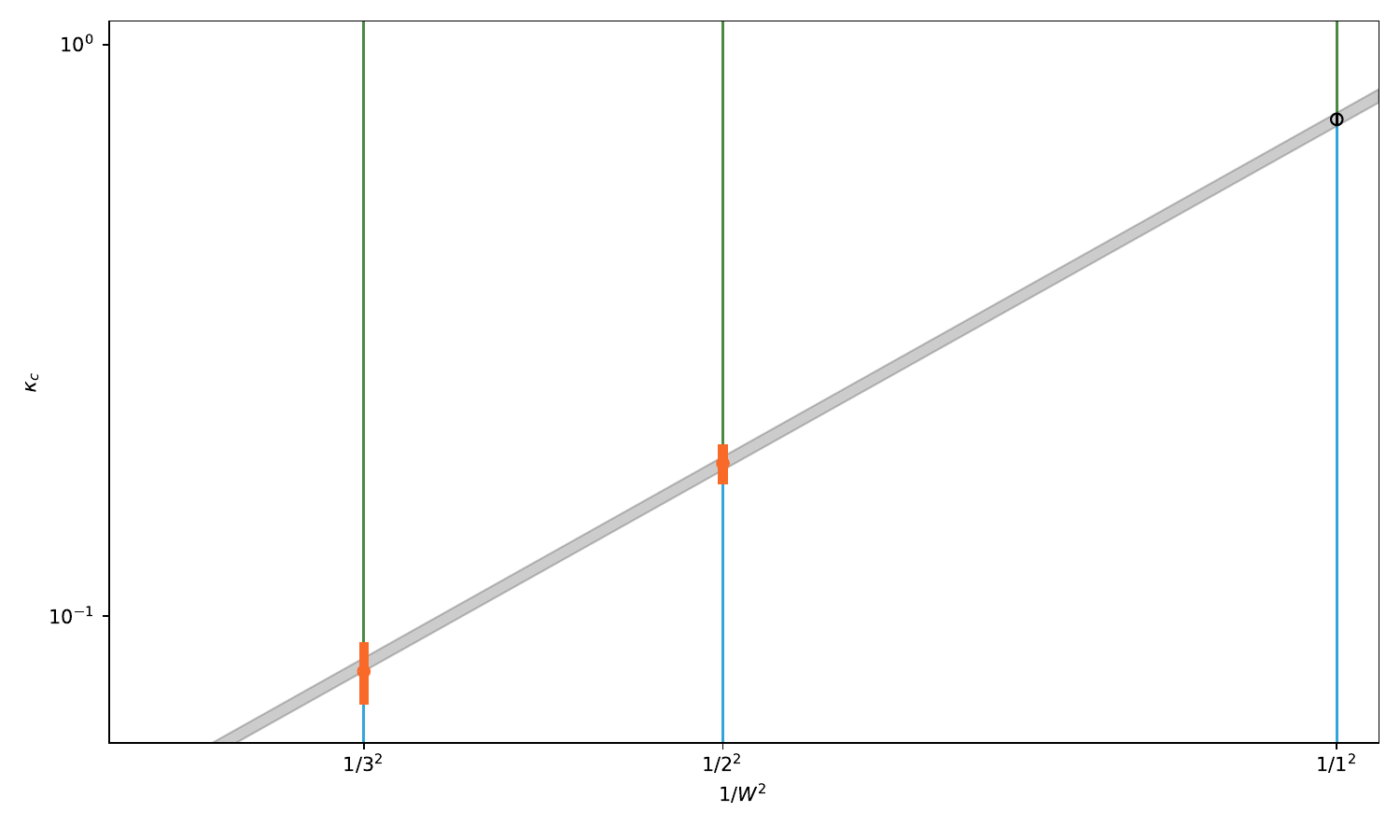}
	\\
	\includegraphics[width=0.49\textwidth]{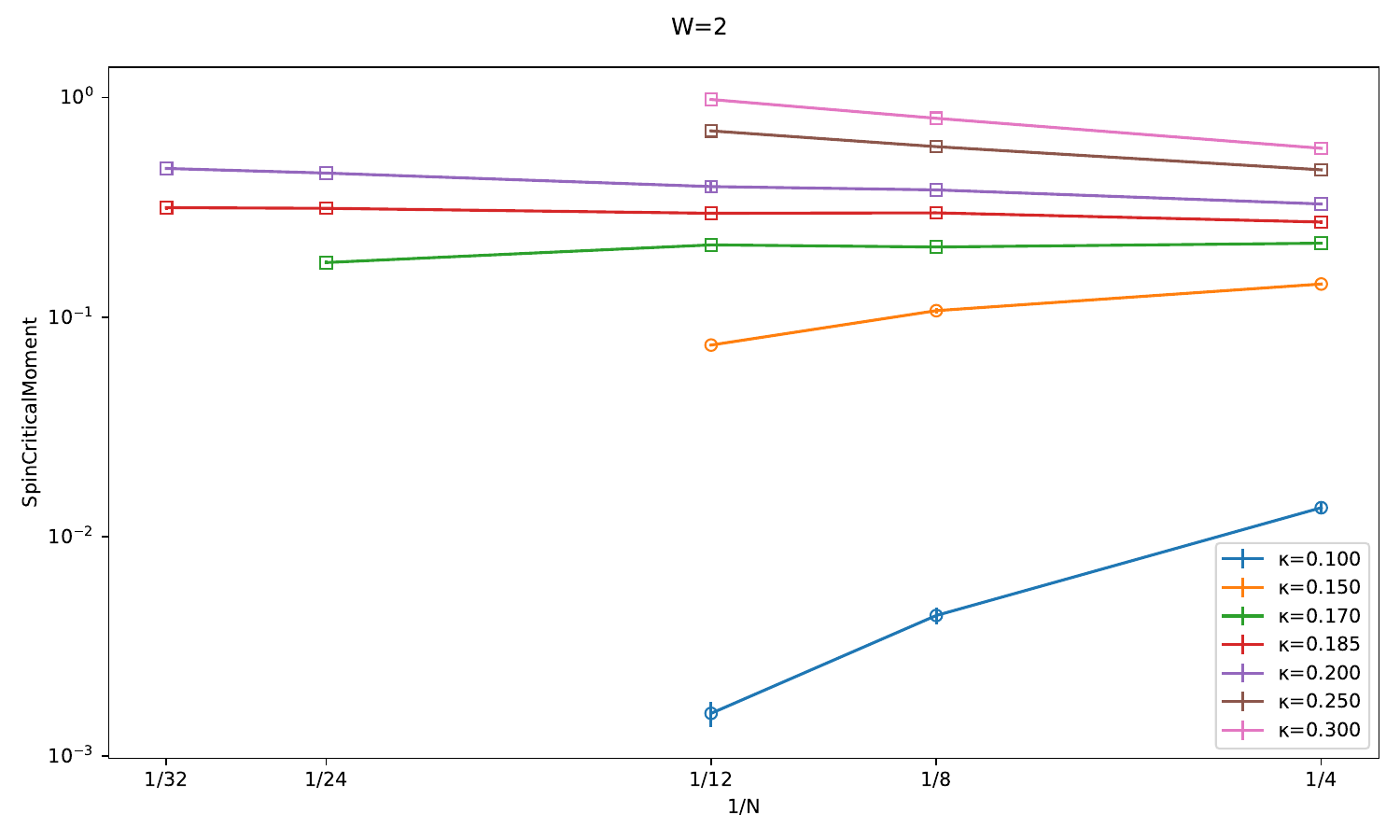}
	\includegraphics[width=0.49\textwidth]{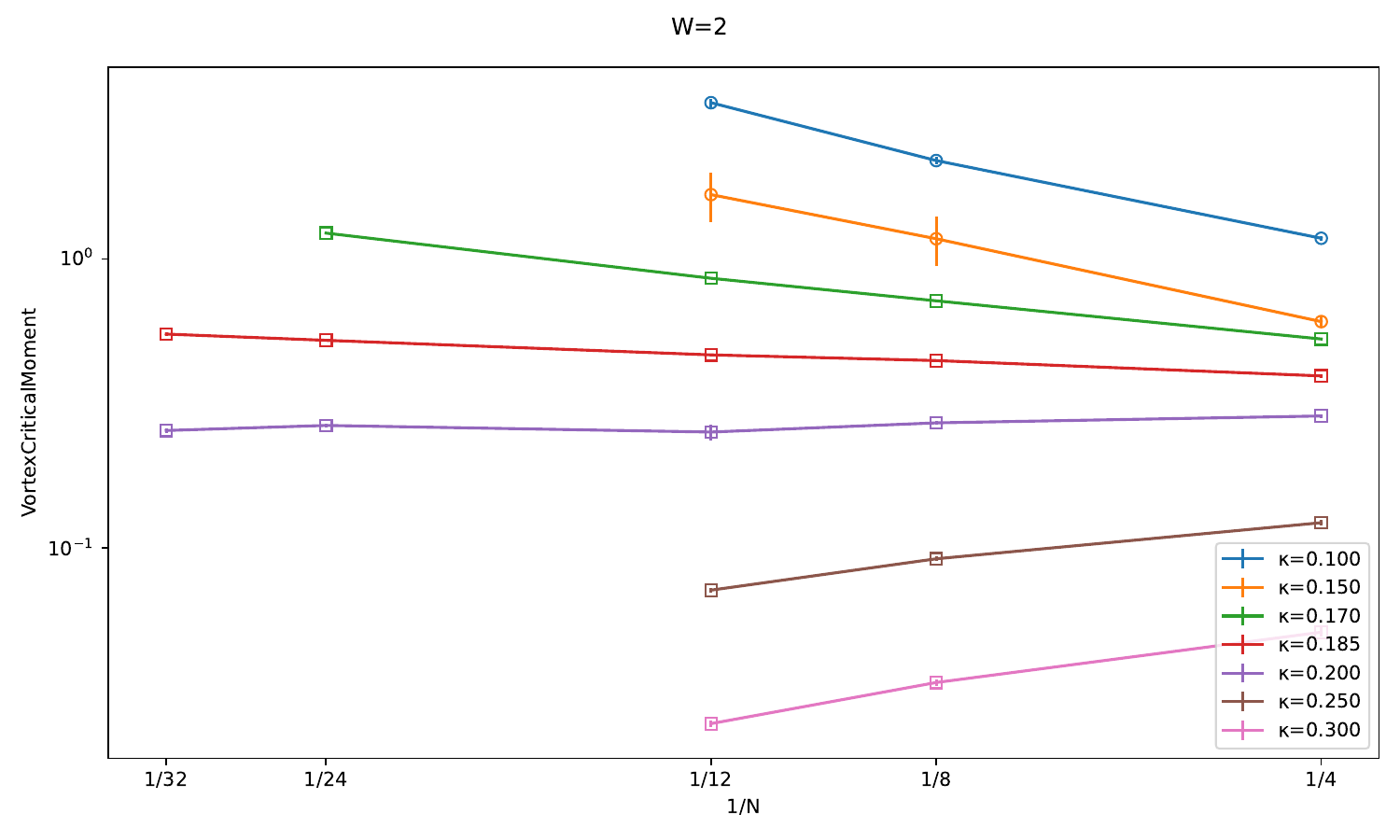}
	\\
	\includegraphics[width=0.49\textwidth]{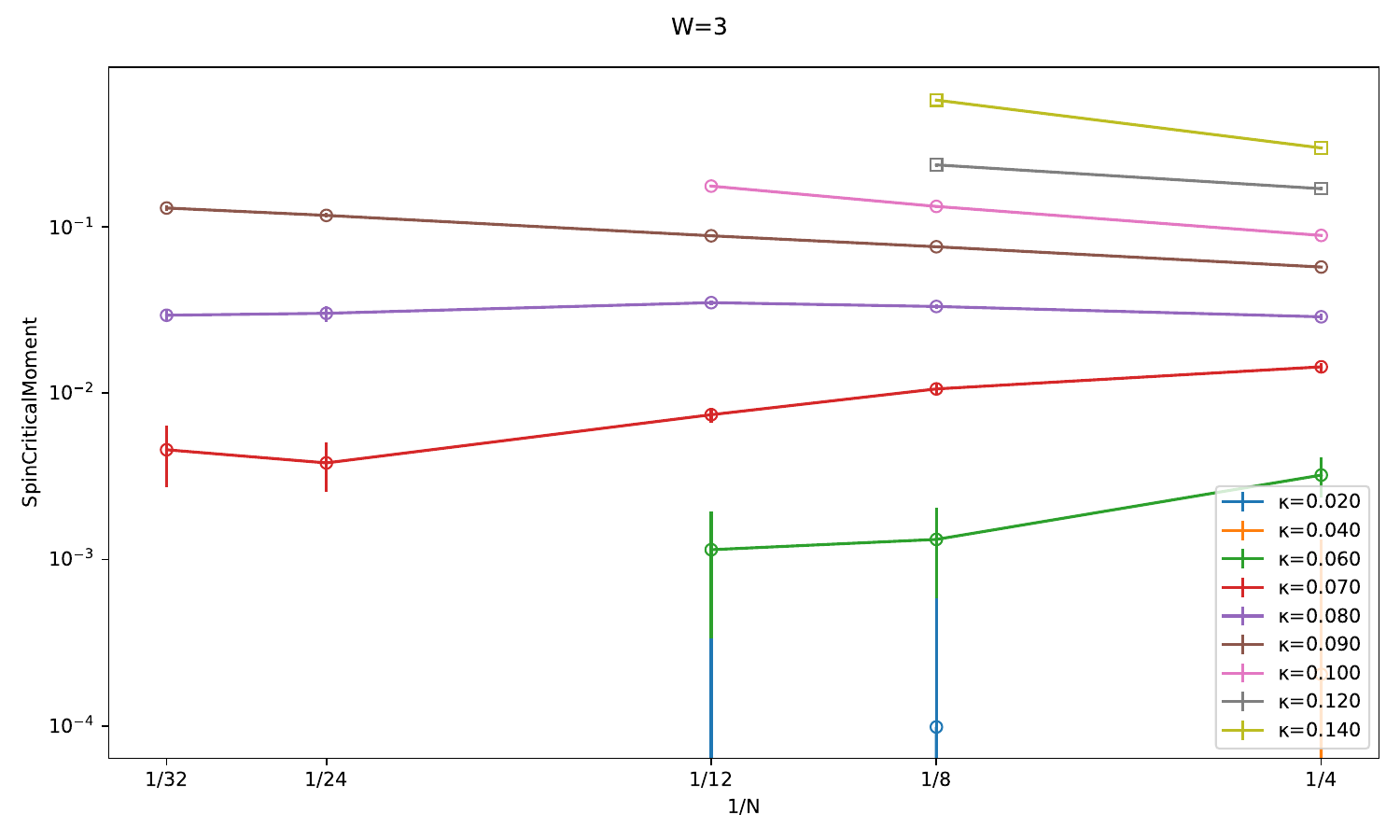}
	\includegraphics[width=0.49\textwidth]{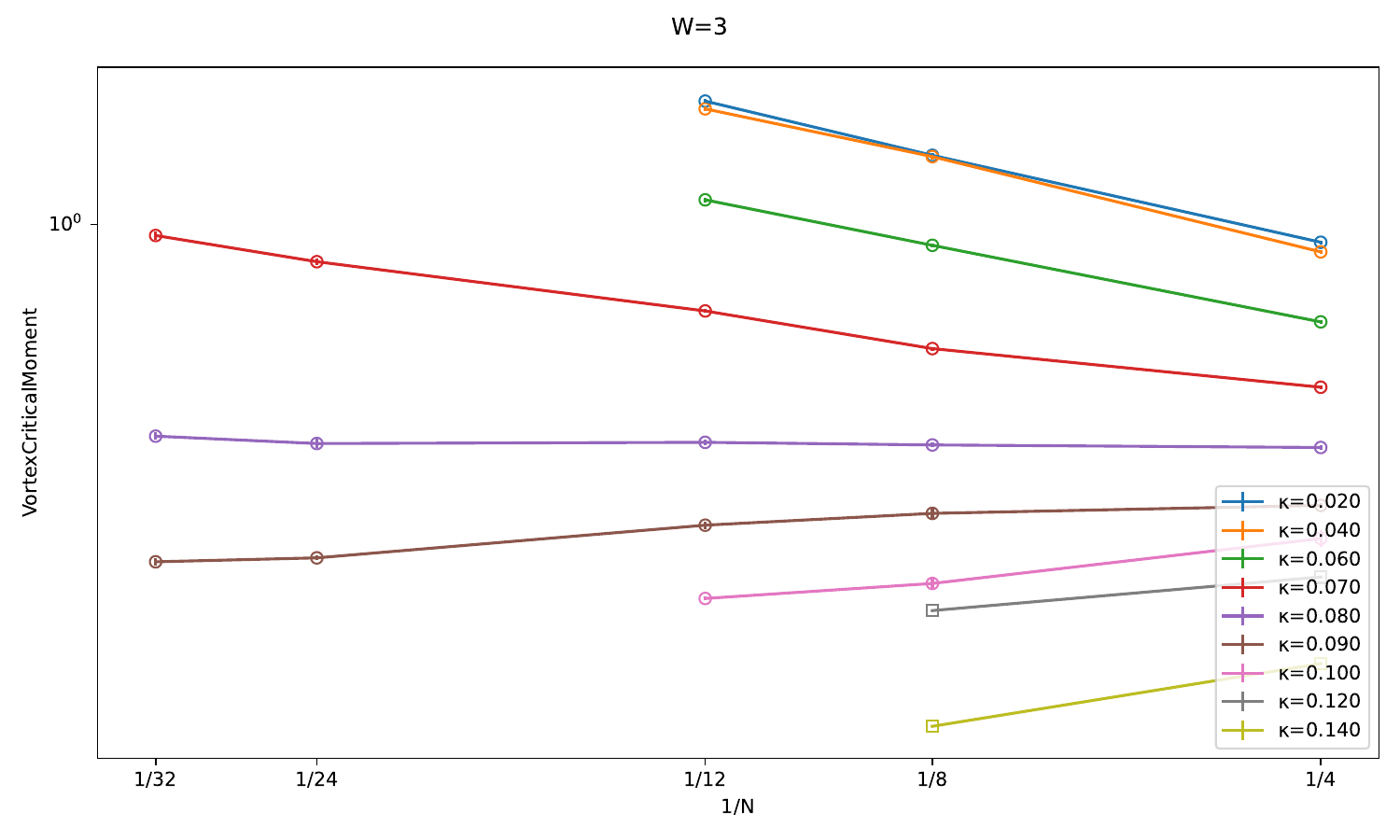}
	\end{center}
	\caption{
		Preliminary numerical results for a variety of $W$, increasing with each row.
		The left column shows the spin critical moment $C_S$ \eqref{spin critical moment}, the right the vortex critical moment $C_V$ \eqref{vortex critical moment}; in both the infinite-volume limit $N\goesto\infinity$ is to the right.
		At the critical coupling the critical moments are flat, while away from the critical coupling the critical moments either go to 0 or diverge.
		When $W>1$ the 't Hooft anomaly guarantees that one of the two diverges in the infinite-volume limit, and we can see that at a fixed $W$ a single coupling $\kappa$ (various colors) are either above or below the nearly-flat (and therefore close-to-critical) $\kappa$.
		Squares indicate a worldline computation, circles indicate a Villain compuation.
		UPPER RIGHT:
		Because the vortex correlator $V$ \eqref{vortex critical moment} is trivial when $W=1$, we instead show the estimated critical coupling $\kappa_c$ for $W=\{1,2,3\}$.
		The transitions are estimated to be at the $\kappa$ which yielded the flattest critical moments at each $W$, with uncertainties given by the mean difference between that $\kappa$ and the closest on either side.
		The black error band is determined entirely by the $W=1$ result (the black point) and the analytic $W^{-2}$ scaling without fitting to the orange points.
	}
	\label{fig:results}
\end{figure}

In \Figref{results} we show the critical moments $C_S$ \eqref{spin critical moment} and $C_V$ \eqref{vortex critical moment} for $W\in\{1,2,3\}$ and a variety of couplings $\kappa$.
We see some $\kappa$ yield near-constant critical moments; these are the critical couplings $\kappa_c$.
Precision critical values are not extremely important, and we estimate them by taking the flattest curve and assigning uncertainties based on the nearest couplings we scanned.

We see that if a coupling is above the flat curve for one critical moment it is below the flat curve for the other, so that we have order of one kind or the other regardless of $\kappa$, as required to satisfy the lattice 't Hooft anomaly.
In the upper right panel of \Figref{results} we check the analytically-expected $W^{-2}$ scaling of the critical coupling and find excellent agreement, as explained in the caption.
We conclude that this lattice action gets the mixed 't Hooft anomaly correct even at finite spacing, and that this guarantees order: the large-coupling regime is the compact boson CFT while the other side has spontaneous symmetry breaking of the winding $\ZZ_W$.
The transition is infinite-order, inherited from the $W=1$ BKT case.
We look forward to $W=\infty$ computations that land on a CFT for any value of $\kappa$ with no fine tuning whatsoever.

\acknowledgments

\href{https://conference.ippp.dur.ac.uk/event/1265/contributions/7595/}{This talk} immediately followed \href{https://conference.ippp.dur.ac.uk/event/1265/contributions/7586/}{the talk of A. Cherman} and relied on that context to elide many details.
We gratefully acknowledge the computing time on the supercomputer JURECA~\cite{jureca-2021} at Forschungszentrum J\"{u}lich through the VSR grant 30278.

\FloatBarrier
\clearpage
\bibliographystyle{JHEP}
\bibliography{master}

\end{document}